\documentclass[anonymous=false,acmsmall,screen,nonacm]{acmart}

\AtBeginDocument{%
  }

\begin{document}

\title{Boosting Visual Fidelity in Driving Simulations through Diffusion Models}

\author{Fanjun Bu}
\affiliation{%
  \institution{Cornell University, Cornell Tech}
  \city{New York City}
  \country{United States}
}
\email{fb266@cornell.edu}

\author{Hiroshi Yasuda}
\affiliation{%
 \institution{Toyota Research Institute}
 \city{Los Altos}
 \country{United States}}
\email{hiroshi.yasuda@tri.global}

\renewcommand{\shortauthors}{F. Bu and H. Yasuda}

\begin{abstract}
Diffusion models have made substantial progress in facilitating image generation and editing. As the technology matures, we see its potential in the context of driving simulations to enhance the simulated experience. In this paper, we explore this potential through the introduction of a novel system designed to boost visual fidelity. Our system, DRIVE (Diffusion-based Realism Improvement for Virtual Environments), leverages a diffusion model pipeline to give a simulated environment a photorealistic view, with the flexibility to be adapted for other applications. We conducted a preliminary user study to assess the system's effectiveness in rendering realistic visuals and supporting participants in performing driving tasks. Our work not only lays the groundwork for future research on the integration of diffusion models in driving simulations but also provides practical guidelines and best practices for their application in this context.
\end{abstract}

\begin{teaserfigure}
  \includegraphics[width=\textwidth]{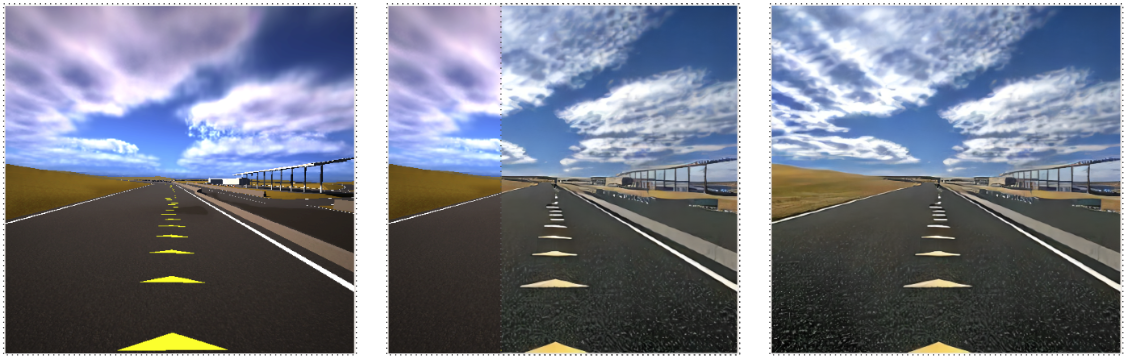}
  \caption{Using diffusion models, we can style transfer the view within a driving simulation into a photorealistic image in near real-time. The leftmost image is taken within a driving simulation in Unreal Engine 5.3, and the rightmost image is the output of our diffusion pipeline.}
  \Description{A panel of three images featuring the same scene within a driving simulation. The image is taken from the view of a driver, sitting in a vehicle on the road. The left image is the original view in the simulation. In the middle image, the right two-thirds of the image is overlayed with the output of a diffusion model which makes it more realistic. The right image is the complete output image from the diffusion model.}
  \label{fig:hero}
\end{teaserfigure}

\received{20 February 2007}
\received[revised]{12 March 2009}
\received[accepted]{5 June 2009}

\maketitle

\section{Introduction}
Driving simulators have been widely used in both academic research and industry due to their cost-effectiveness and safety benefits \cite{blaauw1982driving, kaptein1996driving}. As driving simulations serve as proxies for real-world driving environments, the question of validity is a recurring topic of discussion.
While various definitions of validity exist in the context of driving simulations, physical validity and behavioral validity are the two commonly accepted forms of validity to evaluate the effectiveness of driving simulations \cite{godley2002driving}. Physical validity reflects how the simulator replicates its real-world counterpart, including aspects such as visual fidelity and vehicle dynamics; behavioral validity concerns whether the driving simulator can enable participants to behave as if the simulated situation were real. In our work, we investigate the improvement of physical validity through visual enhancement.

Research has shown that visual fidelity can significantly influence drivers' performance and behavior, particularly in attention-critical applications, as varying levels of visual fidelity may alter drivers' attentional responses \cite{merenda2019effects, zhao2018use, van2015effects}. Consequently, our work aims to enhance the visual fidelity of existing driving simulations, hoping such improvement will lead to increased behavioral validity in downstream user studies.

Traditional methods of increasing visual fidelity in driving simulations involve labor-intensive efforts to create realistic textures and meshes, which also heavily rely on powerful rendering engines \cite{galazka2021cithrus2, zhou2023garchingsim, minhas2016lee, dosovitskiy2017carla}. Such efforts typically require multiple artists specialized in game object design and are challenging to generalize across different simulated environments. However, with advancements in computer graphics technologies, particularly generative models, there is a growing shift from rendering-based methods in 3D space to data-driven approaches in 2D-pixel space \cite{yurtsever2022photorealism, saadatnejad2021shared, li2017photo, chen2021geosim, kim2021drivegan}, which offer better scalability.

In this paper, we explore the use of diffusion models as a data-driven method to improve visual fidelity directly in the image space. Building on previous research, we minimize the rendering effort in the 3D scene and delegate the task of aesthetic enhancement to a data-driven diffusion model. Consequently, we introduce the DRIVE system, an interactive platform that shifts the burden of high-fidelity rendering from the simulation engine to a diffusion model. We demonstrate the capabilities of the DRIVE system through a preliminary user study and discuss its potential for other forms of visual enhancement as well.

\section{Related Works}
\subsection{Photorealistic Driving Simulations}
The traditional approach to achieving photorealistic driving simulations relies on powerful rendering engines, as well as meticulously crafted meshes and textures \cite{dosovitskiy2017carla, shah2018airsim, galazka2021cithrus2}. However, new 3D models and textures must be created when adapting to new environments. For example, in GarchingSim, \citet{zhou2023garchingsim} employs a team of specialized artists to create assets in the Unity3D game engine. To achieve a photorealistic scene in Unity, they also have to implement the High Definition Render Pipeline (HDRP) in Unity. 

To reduce costly human labor, hybrid methods that combine image-based and geometry-based techniques have been proposed. \citet{ono2005photo} designed a system where only nearby objects, typically those users interact with, are rendered using geometry-based methods, while distant objects are rendered from a recorded video image dataset using image-based techniques. With advancements in generative models, the image-based component of this hybrid approach can be replaced with synthesized backgrounds generated by Generative Adversarial Networks (GANs) \cite{yurtsever2022photorealism}, and the generation process is usually conditioned with a semantic map to impose constraints \cite{yurtsever2022photorealism, saadatnejad2021shared}. The synthesis of foreground-rendered objects with background images is typically achieved through dedicated hardware or blending techniques like alpha blending. 

Leveraging internet-scale data, researchers can now simulate dynamic environments purely in pixel space without a backend physics engine. Neural radiance fields (NeRFs) have gained popularity in separately modeling the foreground dynamic objects and background static environments with real-world image/video datasets, enabling the simulation of novel, photorealistic vehicle behavior or changing camera view angles \cite{yang2023emernerf, wu2023mars, tonderski2024neurad}. To address the limitation of NeRFs' slow training and rendering process, Gaussian Splatting techniques are suggested as an alternative for real-time rendering \cite{cheng2024gaussianpro, yan2024street}. Moreover, the data-driven neural simulator proposed by \citet{kim2021drivegan} is trained on recorded driving videos and associated human maneuvers to learn the transition between frames. As such, the simulator can output the next frame purely based on the previous frame and input action, without the help of traditional physics engines such as Unreal Engine and Unity. 

With the help of machine learning and large datasets, we are witnessing a shift from complex 3D rendering to sophisticated 2D image post-processing in the literature of driving simulations.

\subsection{Image Style Transfer}
Once in image space, the task of transforming a 2D image captured in simulation into a photorealistic scene can be approached as a style transfer problem \cite{sohaliya2021evolution, yoo2019photorealistic}. This involves transferring the content of the input image to match the style of a realistic image. Researchers have applied style transfer algorithms in various contexts, from enhancing visual fidelity in surgical training to improving sim-to-real transfer in robot learning \cite{ikeda2022sim2real, luengo2018surreal}. With advancements in generative models, the primary methods for style transfer have evolved from traditional image feature matching to data-driven approaches, such as diffusion models \cite{li2024diffstyler, wang2023stylediffusion, zhang2023inversion, hartley2024domain, xia2021real}. There are multiple techniques for performing style transfer using diffusion models, with some of the simplest methods involving prompt engineering or fine-tuning existing models like Stable Diffusion \cite{everaert2023diffusion}. Style transfer using diffusion models is an active area of research and extends beyond the scope of this paper. In our approach, we employed a parameter-efficient fine-tuning method, Low-Rank Adaptation (LoRa) \cite{hu2021LoRa}, as a proof of concept.

\begin{figure}[h]
    \centering
    \includegraphics[width=\linewidth]{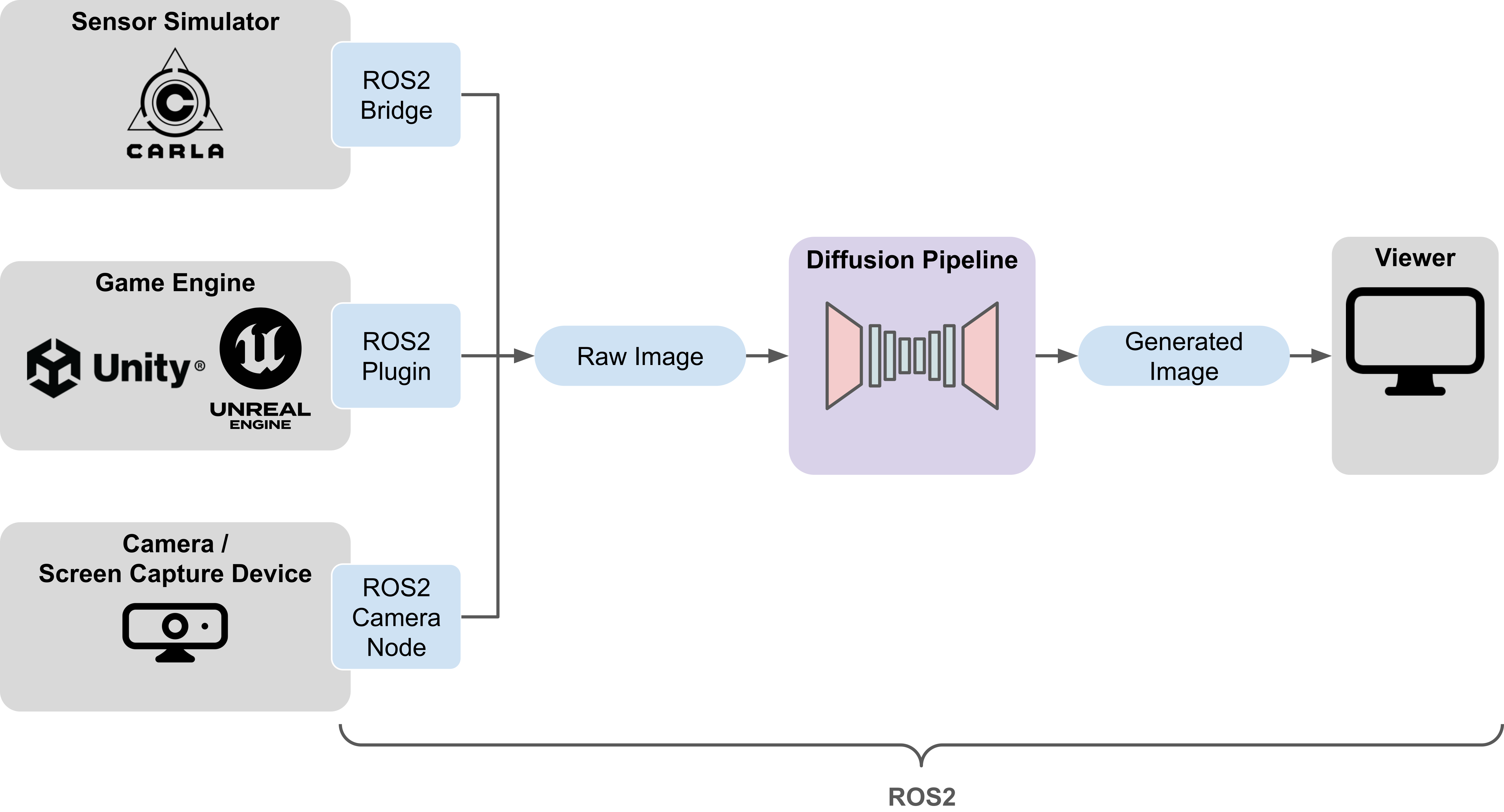}
    \caption{A system diagram for the DRIVE system.}
    \Description{System diagram for the DRIVE system. On the left displays three different sources of image inputs, including Sensor Simulator, Game Engine, and Screen Capture Device/Camera. The inputs are fed to the diffusion pipeline, and the output images are presented to the viewer.}
    \label{fig:DRIVE}
\end{figure}

\section{The DRIVE System}

In this section, we introduce the DRIVE system, which harnesses the power of diffusion models in driving simulations (Fig. \ref{fig:DRIVE}). To ensure the system's generalizability and adaptability across various applications, we used Robot Operating System 2 (ROS 2 Humble \footnote{https://docs.ros.org/en/humble/index.html}) to facilitate the communications among different components within the system. A system diagram is shown in Fig. \ref{fig:DRIVE}. The input raw image data can originate from diverse sources, including pre-existing Sensor Simulators like Carla \cite{dosovitskiy2017carla}, 3D Game Engines like Unreal Engine and Unity \cite{haas2014history, unrealengine}, or screen Capture devices. ROS 2 will fetch images from different modules within the system and broadcast them back as Image type messages. While the system supports diverse sources, the following sections describe the system with Unreal Engine 5.3 as an example.

\subsection{Pipeline Overview}
The core of the DRIVE system is a diffusion model pipeline designed to process image streams in real-time. Our pipeline leverages the existing Stable Diffusion V1.5 framework\footnote{\url{https://huggingface.co/runwayml/stable-diffusion-v1-5}} \cite{rombach2022high}. To optimize computational efficiency and accelerate inference, we have substituted the original variational autoencoder with a Tiny Autoencoder (TAESD)\footnote{\url{https://github.com/madebyollin/taesd}}, a modification commonly adopted within the community. The architecture of our pipeline is illustrated in Figure \ref{fig:pipeline}. The remainder of this section delves into the essential requirements for deploying the diffusion model in driving simulations and discusses our approach to ensuring the model adheres to these constraints.

\subsection{Acceleration}
The diffusion pipeline must process images in real-time to maintain the basic functionality of driving simulations. To this end, we combined multiple approaches to accelerate our pipeline. 

We first looked into different acceleration optimizers that are publicly available, specifically xFormer, TensorRT\footnote{\url{https://github.com/NVIDIA/TensorRT}}, Stable Fast\footnote{\url{https://github.com/chengzeyi/stable-fast}}, and the torch compile function came with PyTorch \cite{xFormers2022, paszke2017automatic}. Our experiments indicate that Stable Fast provides the best performance because it is specifically optimized for Stable Diffusion Pipelines. To further enhance the processing speed, we integrated distilled latent consistency models (LCM) in our pipeline, significantly reducing the number of required inference steps per image \cite{luo2023latent}. Specifically, we injected LCM LoRa adapters in our pipeline, adjusting the pipeline to process images at a resolution of 640 x 480 for only one inference step \cite{luo2023lcm}. Our final effort to enhance the performance of the pipeline involved distributing the computation across multiple machines. Specifically, we used two different machines, one for Unreal Engine and one for the diffusion pipeline, to distribute the computation load. Currently, Unreal Engine operates on a laptop equipped with an i9-9900KF CPU and an Nvidia GeForce RTX 2080 Mobile GPU. Meanwhile, the diffusion model runs on a desktop powered by an i9-11900K CPU and an Nvidia GeForce RTX 3090 GPU. Note that since the aesthetic is now handled by the diffusion pipeline, the 3D rendering in Unreal Engine can be kept as simple as possible; users can reduce rendering quality in the simulation to save computing without affecting the output quality from the diffusion pipeline. The communication between computers is also managed by ROS 2. This division of computational tasks ensures each component of our system is running on dedicated hardware that maximizes its performance potential.

Currently, our system achieves a processing rate of 10 frames per second for an input resolution of 640 x 480, demonstrating its potential as an experimental framework for real-time driving simulation. While we can increase input image resolution, there is a trade-off between image resolution and processing speed. In the context of driving simulation, we prioritize speed over resolution, hence we capped the resolution at 640 x 480.

\begin{figure}[h]
    \centering
    \includegraphics[width=0.8\linewidth]{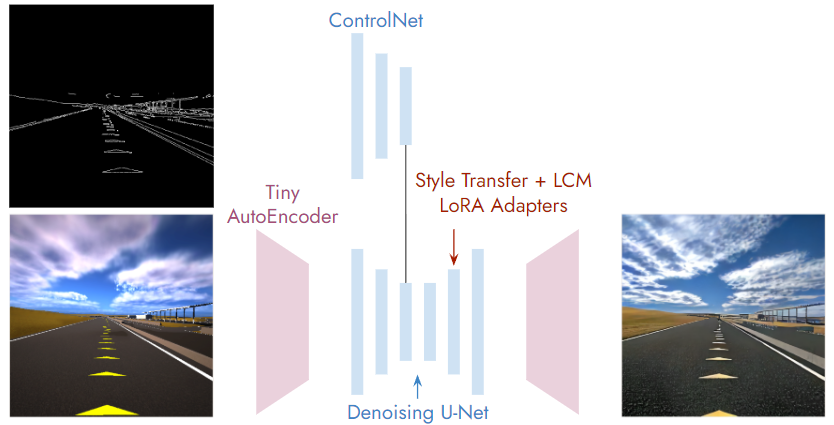}
    \caption{Diffusion model pipeline that gives our simulation a photorealistic view. ControlNet was used to ensure the generative process does not alter the lane marking on the road. LoRa adapters are used to perform style transfer and incorporate latent consistent models. The pipeline is currently running at 10 fps for an input image of resolution 640x480. (For illustration, the images shown above are rendered at a resolution of 512 x 512 with 5 inferencing steps.)}
    \Description{A diagram for the diffusion pipeline. On the left is the input image pair, which includes the original image and its associated edge map. The edge map is fed to ControlNet, while the original input image is processed through the standard stable diffusion pipeline. which includes an image encoder (Tiny AutoEncoder), a denoising U-Net(where the control signal from ControlNet is incorporated), and a decoder to decode the image. The pipeline also features LoRA adapters for style transfer and acceleration through latent consistency models.}
    \label{fig:pipeline}
\end{figure}

\subsection{Within-frame Consistency}
We define \textit{within-frame consistency} as the requirement to keep a selective set of features from the input image invariant in the corresponding output image. In the context of driving, we must guarantee the shape and contour of lane markings on the road remain the same in the output image, which is critical for driving tasks. To address this concern, we leverage ControlNet to condition the generative process with an edge map of the input image (Fig. \ref{fig:pipeline}) \cite{zhang2023adding}. Through ControlNet, we inject spatial conditioning controls into the pre-trained model to maintain the contours invariant between every input image and its associated output image.

\subsection{Cross-frame Stability}
In the context of driving simulations, the information contained within consecutive frames is highly overlapped. We define \textit{cross-frame stability} as the necessity to maintain the overlapping information consistent throughout the frames. For example, if a vehicle is approaching a landmark, the landmark must have the same appearance in all images that capture it. While maintaining such consistency is essential in all driving simulations, it poses a unique challenge for diffusion models, which are inherently generative. Even if the same image is input multiple times, diffusion models may produce vastly different outputs. In a way, the diffusion model behaves like a one-to-many function; however, our goal is to achieve a one-to-one mapping, where the same input reliably produces the same output. 

The variability in outputs is primarily due to the noise introduced during the diffusion process. To counter this, we ensure the noise pattern added to each image remains identical by re-seeding the pipeline with the same seed for every image, thereby eliminating the stochastic nature of the diffusion process and imposing deterministic behavior on the model.

\subsection{Simulation Environment and Photorealistic Style Transfer}
In this paper, we explore a specific application of our diffusion pipeline aimed at enhancing the realism of driving simulations.
Implemented in Unreal Engine (UE) 5.3, our base simulation environment features a simplified Thunderhill Race Track located in Willows, California (Fig. \ref{fig:topdown}), and users can drive and explore the virtual environment freely.

\begin{figure}[h]
    \centering
    \includegraphics[width=\linewidth]{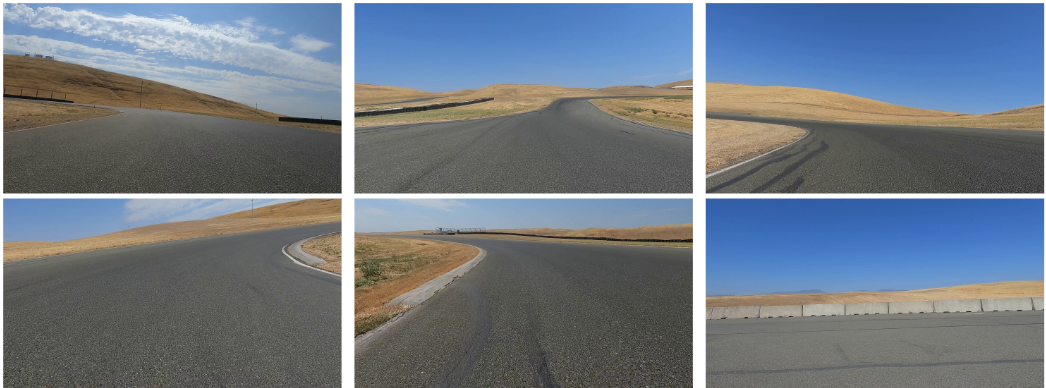}
    \caption{An overview of images in the racetrack dataset for LoRa adapter training.}
    \Description{Six sample images from the race track dataset. Each image is taken from the perspective of a drive on the racetrack, featuring the track surface, the grass on the side, and the blue sky with clouds. }
    \label{fig:dataset}
\end{figure}

To give the simulated environment a realistic view, we performed style transfer using  Low-Rank Adaptation (LoRa), where we trained our own LoRa adapter that represents the realistic Thunderhill Racetrack style\cite{hu2021LoRa}. The training image dataset for this adapter was collected by mounting a camera for video recording on a moving vehicle on the racetrack. The vehicle drives around the racetrack for eight laps, and we sampled 3000 images uniformly from the video covering the entire racetrack (Fig. \ref{fig:dataset}). The images are then used to train a LoRa adapter that captures the aesthetic features of the racetrack.

\section{Preliminary User Study}

To assess the DRIVE system's effectiveness in rendering realistic visuals and supporting participants in driving tasks, we run a preliminary user study where the user needs to perform driving tasks with and without the DRIVE system.  

\subsection{Study Setup}
\begin{figure}[h]
    \centering
    \includegraphics[width=\linewidth]{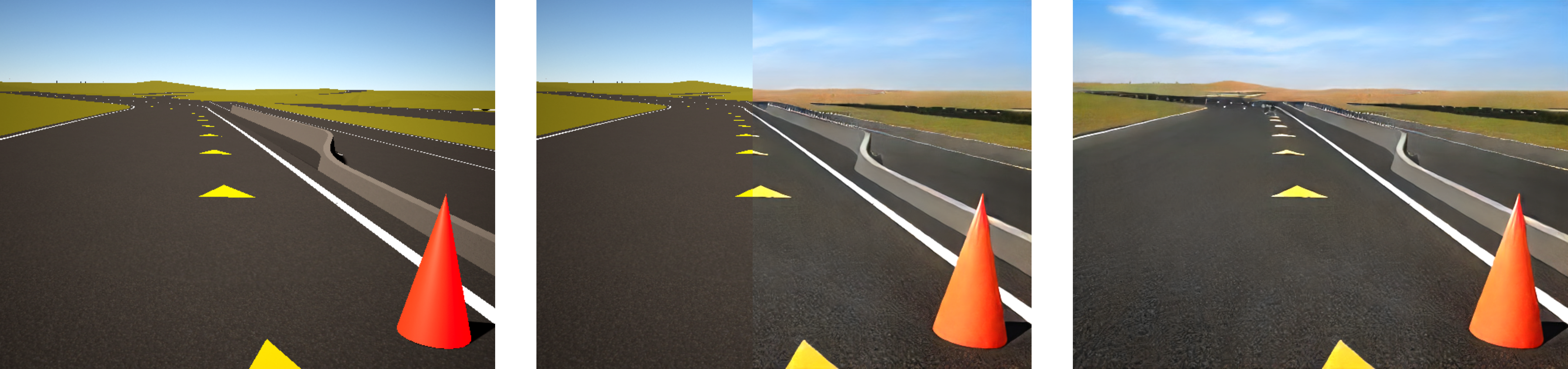}
    \caption{Simulated racetrack environment for our preliminary user study. The leftmost image is the input to the DRIVE system, which is what the participant will see in condition B. The rightmost image is the output image from the DRIVE system, which is what the participant will see in condition A. Since we prioritize inference speed, the diffusion pipeline only takes one inference step for every input image to perform style transfer.}
    \Description{A panel of three images showing the simulated racetrack environment for our user study. Similar to Figure 1, The leftmost image is the input to the DRIVE system, which is what the participant will see in condition B. The rightmost image is the output image from the DRIVE system, which is what the participant will see in condition A. The middle image shows the aforementioned two images side by side to show contrast.}
    \label{fig:pipeline}
\end{figure}
\begin{figure}[h]
    \centering
    \begin{minipage}[b]{0.49\textwidth}
        \centering
        \includegraphics[height=0.6\textwidth]{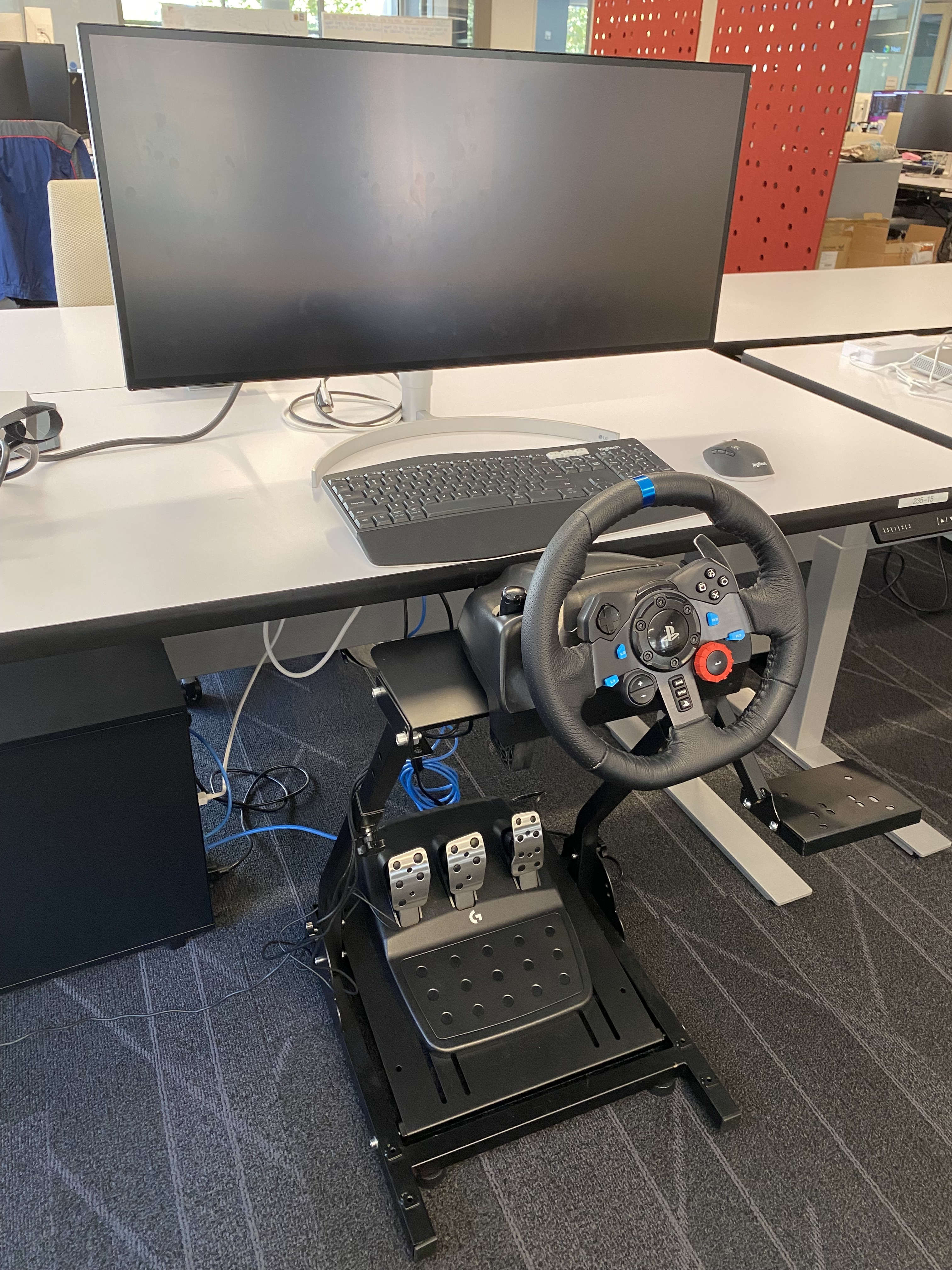} 
        \caption{The hardware setup for our user study.}
        \Description{Hardware setup for the user study. The description can be found in the paper.}
        \label{fig:study}
    \end{minipage}
    \hfill 
    \begin{minipage}[b]{0.49\textwidth}
        \centering
        \includegraphics[height=0.6\textwidth]{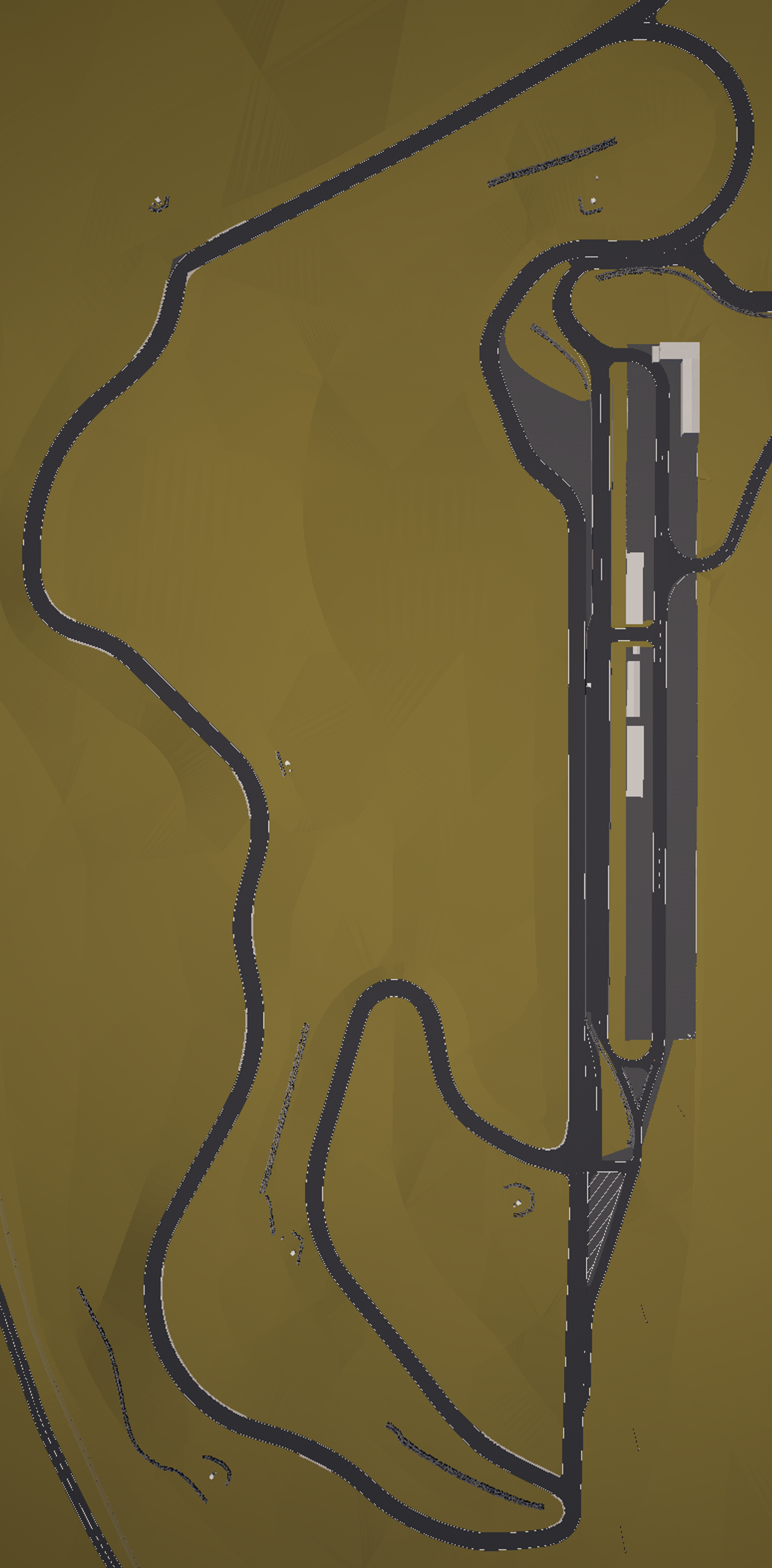} 
        \caption{A top-down view of the simulated racetrack environment. The participants drive counter-clockwise.}
        \Description{A top-down view of the simulated racetrack.}
        \label{fig:topdown}
    \end{minipage}
\end{figure}

Our study takes place in a standard screen-based fixed-based driving simulator. A set of Logitech G29 Driving Force Racing Wheel and Floor Pedals are mounted on a rigid frame, and stationed in front of a flat monitor screen (Fig. \ref{fig:study}). The participants will sit in front of the steering wheel and operate the simulator as if they are in a real vehicle. For pedals, the right pedal is the acceleration pedal, the middle pedal is for the brake. The left pedal is not used in this experiment. 

We choose motorsport racing as the study context because it provides a simplistic, relatively static environment, allowing participants to focus on the visual perspective of the simulation. 
Specifically, the participant will drive on the two-mile track counterclockwise. The racetrack is overlayed with the racing line in the form of a sequence of yellow arrows on the floor. The racing line is the optimal vehicle trajectory that would achieve the best possible lap time\cite{botta2012evolving}, and is collected from an expert driver's demonstration. The participant's task is to drive around the racetrack following these arrows for two laps. Since the frame rate is only 10 frames per second, we keep the maximum driving speed at 50 km/h. 

\subsection{Study Protocol}
Upon arrival, the participants are asked to sign the informed consent. Then, the participants are introduced to our hardware setup. The study contains two conditions. In condition A, the participants are asked to drive two laps following the racing line in the simulated environment with the DRIVE system performing real-time style transfer. In condition B, the participants will perform the same task but see the original images without the DRIVE system enhancement. The condition order is counter-balanced. In both conditions, the simulation is running at 10 frames per second. Upon finishing both conditions, the participants will complete a questionnaire evaluating their experience in both conditions.

\subsection{Participants}
In total, we recruited 9 participants (6 male, 3 female). All participants have a valid driver's license, and one participant has experience with motorsport racing. The participants were recruited based on convenience sampling within the institution, and the study has obtained IRB approval.

\section{Results}

\begin{figure}[h]
\centering
\begin{tabular}{ccc}
  \includegraphics[width=0.3\linewidth]{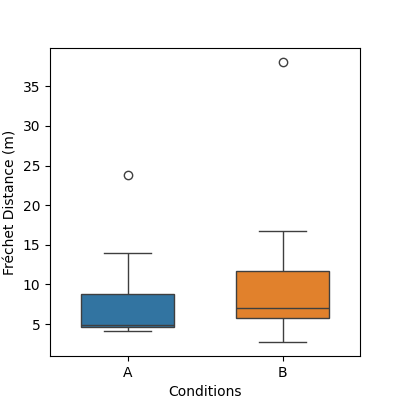} &   \includegraphics[width=0.29\linewidth]{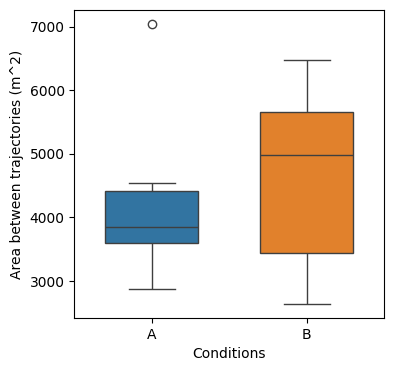} &
  \includegraphics[width=0.29\linewidth]{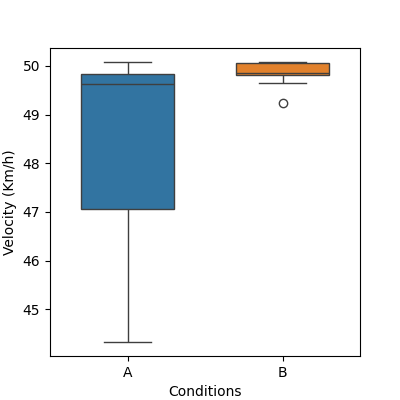} \\
\footnotesize{(a) Fréchet Distance.}  & \footnotesize{(b) Area between racing line and user trajectory.} & \footnotesize{(c) Mean vehicle speed.}
\end{tabular}
\caption{Subjective performance measures of the driving task.}
\Description{Box plots that compare participants' task performance under both conditions. Details can be found in the paper.}
\label{fig:result}
\vspace{-0.4cm}
\end{figure}
\subsection{Task Performance}
To evaluate the users' task performances, we first look at how well their driven trajectories match the racing line. We computed two measures: Fréchet distance and the area between trajectories \cite{eiter1994computing, jekel2019similarity}. 
Fréchet distance can be thought of as a measure of maximum deviation from the racing line, and the area between curves can be interpreted as the cumulative error. While each participant drives two laps under each condition, we only consider the data collected during the second lap and treat the first lap as a warmup lap. As shown in Fig. \ref{fig:result} (a) and (b), both measures suggest that the users perform better under condition A, which is the condition that uses the DRIVE system. Under Condition A, the mean Fréchet distance is 8.23 meters with a standard deviation of 6.28. Under Condition B, the mean Fréchet distance is 11.23 with a standard deviation of 10.33. The mean cumulative area error under Condition A is 4152.6 with a standard deviation of 1128.1. The cumulative area error under Condition B is 4606.4 with a standard deviation of 1308.4. 

While it seems that participants perform better under Condition A, we do not think this provides strong evidence that the DRIVE system improves users' task performance; it only demonstrates that participants are able to perform the task. As the analysis of vehicle velocity shows, participants drive slower with higher variance in speed under Condition A (mean=48.43 km/h, std=1.93) than in Condition B (mean=49.84 km/h, std=0.25). The high performance error under Condition B might be caused by the fact that participants feel more comfortable driving faster in Condition B, suggesting a higher functional fidelity. 

Therefore, from the task performance measures, we can only conclude that the DRIVE system allows participants to perform the same task as with the original simulator. It does not provide a clear picture of how the DRIVE system influences the participants' task performance. 

\subsection{Analysis of Questionnaire Results}
Beyond numerical values collected during the task, we also analyze participants' feedback in the questionnaire, which asks participant's options about each condition in visual fidelity, which is regarding how photorealistic the simulation is, and functional fidelity, which is regarding how well the system supports them to perform the task in terms of maneuverability. 

\begin{figure}[h]
    \centering
    \begin{minipage}[b]{0.49\textwidth}
        \centering
        \includegraphics[width=\textwidth]{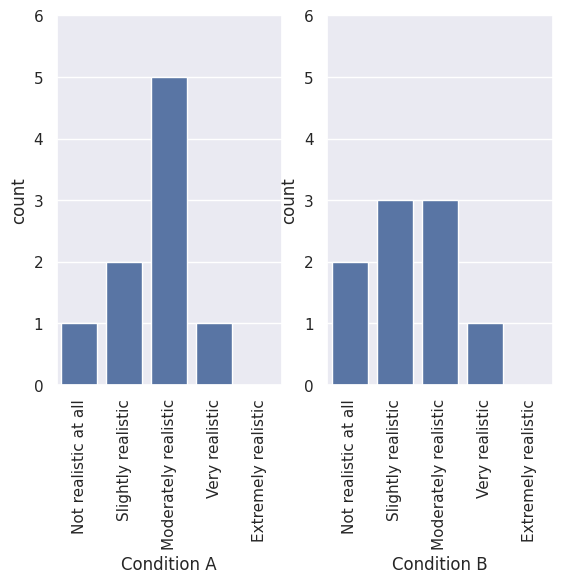} 
        \caption{Users' rating on \textbf{visual fidelity}.}
        \Description{Histograms that compare participants' evaluation of both conditions' visual fidelity on a five-point Likert scale. Details can be found in the paper.}
        \label{fig:figure1}
    \end{minipage}
    \hfill 
    \begin{minipage}[b]{0.49\textwidth}
        \centering
        \includegraphics[width=\textwidth]{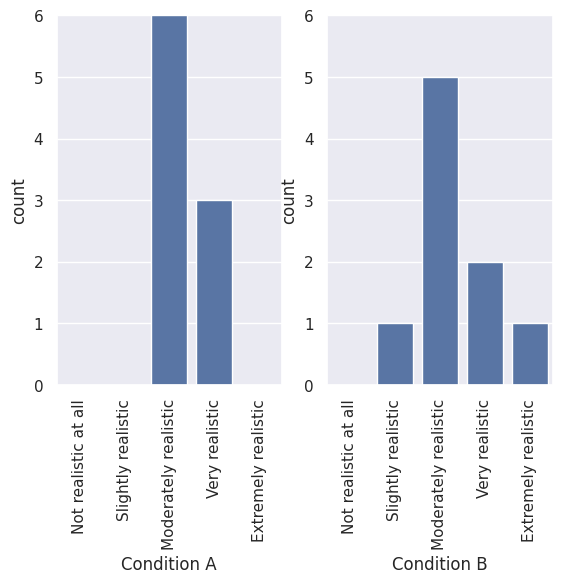} 
        \caption{Users' rating on \textbf{functional fidelity}.}
        \Description{Histograms that compare participants' evaluation of both conditions' visual fidelity on a five-point Likert scale. Details can be found in the paper.}
        \label{fig:figure2}
    \end{minipage}
\end{figure}

When we ask the participants to rate the visual realism of both conditions on a five-point Likert scale, with one being not realistic at all and five being extremely realistic, the mean rating for Condition A is 2.7, which is higher than that for Condition B, which is 2.3. This indicates that the participants think the DRIVE system does improve the visual fidelity of the simulation. Furthermore, when shown screenshots from each condition side by side, eight out of nine participants think Condition A is more photorealistic than Condition B. 

We also ask the participants to rate the functional realism of both conditions on the same five-point Likert scale, which pertains to the maneuverability of the vehicle in the system and how well the system functionally supports them to complete the driving task. Both conditions have an average rating of 3.3, an indication that the DRIVE system does not improve or hinder the simulation's original functionality in the current context.  

When we ask the participants to write down their experience under Condition A (with the DRIVE system), many report that the visual fidelity is better. P1 says, \textit{"[it has] detailed sky and ground textures, smooth edges to objects."} P4 also says, \textit{"Condition B was not photorealistic compared to Condition A, as it lacked shadows or the visual details associated [with] roads, grasses, etc."} P8 reports, \textit{"[Condition A is] more realistic than B, particularly the lighting and extra detail in grass and sky."} P6, the only participant who thinks Condition B is more realistic, says \textit{"[Condition A] feels like a 2D image, so no cue of depth." }
When it comes to Condition B, participants report that it is more game-like. P1 says, \textit{"[Condition B] is blocky, jagged, [and] lack of detail."} P8 comments, \textit{"The visual is flat without much texture. Looks like a basic video game."} P9 reports, \textit{"[Condition B] is less photorealistic but more high contrast."}

However, some participants do report that Condition A is harder to drive within. P4 says, \textit{"While Condition A was more photorealistic, Condition B was easier to drive because the color contrast was clearer, making it easier to distinguish the objects, especially those in distant locations."} P8 also says, \textit{"Condition A was more realistic but I also got distracted that the yellow arrows were flickering between white and yellow."} P5 comments, \textit{"It [kind of] looks like generative AI with the way each frame was coming into [the] picture. It did look more realistic. The driving seemed harder."} On the other hand, P7 says that \textit{"[in Condition B], [the] yellow marker is constant and straightforward ... I feel I am driving in a simulator but with clear instruction from the marker."} P9 reports, \textit{"The increased contrast with condition B (while not being as photorealistic) gave me more confidence in what I was seeing as a driver."} These comments may provide an explanation of why the average driving speed under Condition A is slower than that under Condition B, hinting at a slight drop in functional fidelity under Condition A. 

\section{Limitations}
From the participants' qualitative feedback, we notice that while the DRIVE system increases the visual fidelity of the simulation, the diffusion process may introduce factors that lead to a decrease in functional and behavioral fidelity. We discuss some of the potential reasons below and propose solutions for future works. 

\subsection{Edge Detection with Low-Resolution Input}
Since the image feed coming from the simulator is in relatively low resolution, the output of edge detection can be noisy, especially for faraway objects. Unfortunately, diffusion models are highly sensitive to noise in the input control(edge) image. Despite removing stochasticity from the diffusion process, we have observed that even minor perturbations within the edge map can result in unwanted artifacts, especially around the vanishing point of the road. This is aligned with multiple participants' comments. For instance, P4's reports, \textit{"While [Condition A] seemed more realistic compared to Condition B in terms of color contrast, it made it more difficult to distinguish between the yellow arrows and white line, especially from when they were in further locations, making it difficult to drive."} The edge detection algorithm may also struggle with complex shapes in the input images, such as clouds and signal towers, especially when they approach the viewer at high speed. For the current study, we have excluded these complex shapes to mitigate such issues. In future work, given we have access to the world model and vehicle dynamics model, we will utilize this knowledge to stabilize consecutive edge maps and achieve more consistent control over the generated image sequences.

\subsection{Perception of Optical Flow}
Optical flow is an important source for the perception of motion and distance \cite{lappe2000perception}. In our experiment, since the race track contains long straight portions that do not induce significant scenery changes, the low-resolution simulation environment may not offer the system enough visual details (e.g. asphalt patterns) to support the perception of optical flow. As a result, this may cause the delusion that the vehicle is not moving. This is an inherent problem of the chosen simplistic environment limited by the current system, but we are optimistic that future systems can process more complex input images to support the perception of optical flow. 

\subsection{Inference Speed}
The current system operates at 10 frames per second with a relatively low image resolution, suggesting considerable potential for improvement. One promising direction is the parallelization of the iterative diffusion process, as exemplified by algorithms such as StreamDiffusion and StreamV2V \cite{kodaira2023streamdiffusion, liang2024looking}. Currently, these parallel approaches do not support the integration of additional control signals, such as those provided by ControlNet. Consequently, this approach is not adopted in our existing pipeline configuration. Nonetheless, we anticipate the future inclusion of this capability, which could significantly expedite the diffusion process.

\section{Discussion}
\subsection{Leveraging Motorsport Racing Environments as Research Testbeds}
Motorsport Racing has been one of the popular contexts to collect human behavioral data \cite{shi2021virtual, remonda2021comparing, hisham2017racer, granato2020empirical}. The nature of racing pushes drivers to their performance limits, yet the environment itself remains relatively straightforward compared to other driving scenarios. The simplicity of the environment, combined with clear, objective performance metrics, makes it ideal for user studies. In our research, where the focus is on enhancing visual fidelity, the racing environment provides a controlled test bed, allowing participants to concentrate on visual aesthetics without the distractions of more intricate scenarios.

\subsection{Future Applications}
Although diffusion models have not yet reached a level of maturity sufficient for deployment in safety-critical applications, our pipeline has demonstrated promising potential. Moreover, we believe our pipeline can be generalized for a variety of functionalities in the driving contexts.

\subsubsection{Image Harmonization in Mixed Reality Applications}
Due to the limited realism of motion fidelity in fixed-base simulators, many simulation platforms have transitioned to on-road environments, employing mixed reality technologies to superimpose virtual objects onto real-world views  \cite{bu2024portobello, goedicke2022xr, baltodano2015rrads, ghiurau2020arcar}. A prevalent challenge in these platforms is the visual disparity between foreground objects (virtual) and the background (real world), in terms of texture, visual noise patterns, color profiles, etc.  Such discrepancies often cause the virtual objects to visually dominate, inadvertently drawing the viewer’s attention.
While attention focusing on foreground objects can be beneficial at times, it may counteract the objectives of studies aiming to create a cohesive mixed reality experience. Thus, it can be beneficial to blend the foreground image with the background image, an approach that is commonly defined as image composition or image harmonization in the computer graphics and vision literature \cite{niu2021making}. Recent research has investigated data-driven deep-learning approaches for image harmonization tasks, with diffusion models emerging as a promising new technique. \cite{tsai2017deep, guo2021image, zhou2024diffharmony, li2023image, lu2023painterly}.

\begin{figure}
    \centering
    \includegraphics[width=\linewidth]{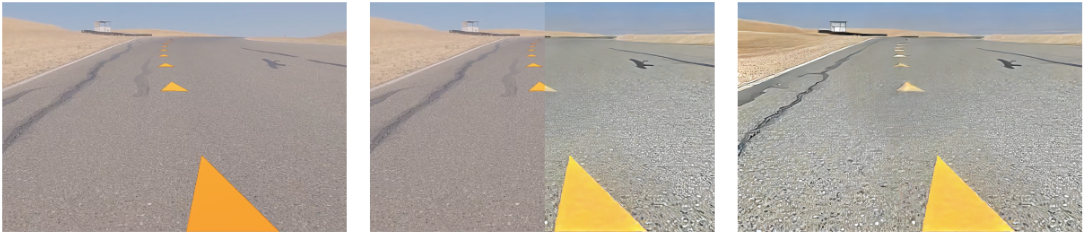}
    \caption{Image harmonization for video pass-through applications.}
    \Description{Demonstration of how our diffusion pipeline can be used for image harmonization. In a panel of three images, the leftmost image shows the original image, which is produced by overlaying virtual racing lines (in the form of a sequence of yellow arrows) on top of recorded camera footage from the racetrack. The middle image shows the original image and the output image from our diffusion pipeline side by side for better contrast. The rightmost image is the output image from our diffusion pipeline, showing a more harmonized view.}
    \label{fig:harmonization}
\end{figure}

We posit that our pipeline is capable of performing real-time image harmonization in driving contexts. As shown in Fig. \ref{fig:harmonization}, we superimpose a virtual racing line onto video-passthrough camera footage and perform style transfer on the entire scene to achieve image harmonization. 

\subsubsection{Interactive Conditioning with Human Feedback}
During our study, we observed that the generated image is highly sensitive to minor variations in the condition image (edge image). This sensitivity provides the possibility of integrating human interactivity into the system. ControlNet offers an intuitive method to guide the generation process of diffusion models and allows for deliberate manipulation of the control image (edge map in our demo) based on human inputs, such as gaze or mouse cursor movements. For example, we could dynamically adjust the edge detection threshold according to the user's gaze, refining the edge map at the focal point and coarsening it peripherally, thereby customizing spatial control in response to the user’s attention. This approach could significantly enhance interactive capabilities, providing more targeted and effective manipulation based on real-time user feedback. 

\section{Conclusion}
In this paper, we introduce DRIVE, a system designed to enhance the realism of virtual environments using diffusion-based techniques. We outline a detailed approach to addressing the challenges of acceleration and consistency in the context of driving simulations. Our preliminary results demonstrate that while DRIVE improves visual fidelity, there remains significant room for improvement in consistency control to better capture scene details. Nevertheless, all participants were able to complete the same driving tasks with and without the DRIVE system, with the majority expressing appreciation for the enhanced visual fidelity. We encourage further development and refinement of our proof-of-concept system from the community to overcome these limitations and unlock its full potential.

\begin{acks}
We thank Sheryl Chau, Hieu Nguyen, and William Kettle, for their support with hardware and software. 
\end{acks}

\bibliographystyle{ACM-Reference-Format}
\bibliography{sample-base}

\end{document}